\documentclass{article}

\usepackage[preprint]{neurips_2026}

\pdfoutput=1

\usepackage[utf8]{inputenc}
\usepackage[T1]{fontenc}

\usepackage{hyperref}
\usepackage{url}
\usepackage{colortbl}
\usepackage{booktabs}
\usepackage{amsfonts}
\usepackage{amsmath}
\usepackage{amssymb}
\usepackage{nicefrac}
\usepackage{multirow}
\usepackage{newtxtext}
\usepackage{textgreek}
\usepackage[LGR,T1]{fontenc}
\usepackage{microtype}
\usepackage{xcolor}
\usepackage{graphicx}
\usepackage{enumitem}
\usepackage{fancyhdr}

\title{ORBIT: Learning Gene Program Co-Activation Structure for Cell-Type-Stratified Pathway Rewiring Analysis in Single-Cell Transcriptomics}

\author{%
\textbf{Yuechen Wang$^{1}$, Lina Jia$^{1}$, Qinglong Wang$^{1}$, and Feng Tian$^{1,*}$}\\[0.5em]
$^{1}$Department of Neurology, Beth Israel Deaconess Medical Center, Harvard Medical School,\\
Boston, MA 02115, USA\\[0.5em]
$^{*}$Correspondence: \texttt{ftian@bidmc.harvard.edu}
}

\begin{document}

\maketitle

\begin{abstract}
Gene programs co-activate within cells, but existing single-cell methods either treat programs independently or require experimental perturbation data to model their interactions.
We introduce ORBIT, a self-supervised transformer that learns asymmetric dependencies among gene programs from observational single-cell RNA-sequencing data alone, quantifying how strongly each program influences every other program.
The key mechanism is an \emph{intervention-consistent} training objective: the model learns each program's directional influence on every other program by predicting how the others change when that program is removed, yielding attention weights that reflect asymmetric influence rather than symmetric co-occurrence.
Applied to 191{,}890 prefrontal cortex nuclei across three pathway vocabularies, ORBIT recovers co-activation structure consistent with established Alzheimer's disease vulnerability signatures, identifies cell-type-specific rewiring invisible to differential expression, and achieves 0.984 macro F1 on cell-type classification from 220 pathway scores, which is within 0.3 points of a state-of-the-art classifier using all 22{,}088 genes. Code is available at: https://anonymous.4open.science/r/elliptic6621.
\end{abstract}

\section{Introduction}

Learning directed dependencies among gene programs from observational single-cell transcriptomes is a central challenge in computational biology~\citep{mou2019,subramanian2005,bull2024}. Gene programs, or curated sets of functionally related genes, co-activate within cells according to structured dependencies that shape cellular phenotype, and these dependencies can be reorganized in disease without any individual program changing its mean expression level~\citep{talukdar2023,seguella2021}. Detecting such \emph{co-activation rewiring} requires a method that (i)~operates on standard single-cell RNA-sequencing data without perturbation labels, (ii)~models pairwise program interactions rather than scoring programs independently, and (iii)~recovers asymmetric influence relationships, not merely symmetric co-occurrence, which is a distinction that determines which program is the upstream driver and therefore the more effective therapeutic target~\citep{kramer2013}. No existing method satisfies all three requirements. In the brain, where cell types maintain tight coupling between synaptic, metabolic, and immune programs, such relational reorganization may distinguish early neurodegeneration from healthy aging~\citep{mitra2024}.

Existing single-cell methods each address a subset of these requirements. \emph{Gene-set and pathway-scoring methods}~\citep{aibar2017,subramanian2005} quantify program activity at single-cell resolution but treat programs as independent features, satisfying none of (ii) or (iii). \emph{Co-expression and gene regulatory network methods}~\citep{aibar2017,xu2024,silkwood2024} recover coordinated molecular modules but typically operate at the gene or transcription-factor level, rely on symmetric associations, and do not directly yield an interpretable program-by-program dependency matrix stratifiable by cell type and condition, satisfying (ii) partially but not (iii). \emph{Perturbation-response models}~\citep{roohani2024,gonzalez2025} learn directed effects and satisfy (ii)--(iii), but require matched perturbation--expression pairs unavailable in most human post-mortem tissue atlases~\citep{mathys2024singlecell,morabito2021}, failing (i). \emph{Single-cell foundation models}~\citep{cui2024,yang2022,theodoris2023} learn powerful cell representations satisfying (i), but their embeddings do not expose an interpretable matrix of pathway-level dependencies comparable across biological states.

We present ORBIT (\textbf{O}rganization of \textbf{R}ewiring \textbf{B}etween \textbf{I}nteracting \textbf{T}ranscriptional programs), a two-stage framework that fills this gap. The idea is simple: train a transformer over program-level embeddings, then add a training objective that forces the attention matrix to encode \emph{directional} dependency. Concretely, for each program $p$, ORBIT zeros $p$'s input and requires the attention weights $A[p,\cdot]$ to predict how every other program's reconstructed score shifts, converting attention from a co-occurrence statistic into a directional dependency operator. Stage~1 learns the co-activation matrix without any labels; Stage~2 fine-tunes a thin classification head ($3.9\%$ of parameters) to stratify the map by cell type and condition. Throughout, we use \emph{directed dependency} to mean asymmetric predictive influence between programs under input ablation, $A[p,q] \neq A[q,p]$ in expectation; we reserve \emph{co-activation} for the underlying expression-covariance structure that ORBIT learns from.

Applied to 191{,}890 prefrontal cortex nuclei from an AD atlas~\citep{morabito2021} across three independent pathway vocabularies, ORBIT recovers biologically interpretable co-activation structure and identifies cell-type-specific rewiring that is structurally undetectable by differential expression. As a secondary validation, the learned pathway representation preserves cell-type identity, achieving classification performance comparable to gene-level cell-type classifiers despite using compact pathway-level inputs.

The main contributions are summarized as follows:
\begin{itemize}[leftmargin=*, itemsep=2pt] 
\item We formally characterize two structural limitations of existing approaches: (1) correlation-based methods are symmetric by construction and cannot represent directed program influence; (2) graph-based perturbation models require experimental labels unavailable in post-mortem tissue, restricting discovery to interactions already encoded in a predefined network.

\item We propose ORBIT, a two-stage framework in which a pathway attention transformer learns asymmetric, directed program co-activation weights from expression covariance alone via an intervention-consistent influence loss, which trains $A[p, \cdot]$ to predict score-space shifts under single-program removal, converting attention from a similarity measure into a directional dependency operator.

\item Applied to an AD snRNA-seq atlas across three pathway vocabularies, ORBIT identifies cell-type-specific rewiring undetectable by differential expression that is consistent with prior empirical evidence and achieves 0.984 macro F1 on cell-type classification from 220 pathway scores (within 0.3 points of CellTypist using all 22,088 genes).
\end{itemize} 

\section{Method}
\label{sec:method}

\textbf{Overview.}
ORBIT takes as input a nucleus's gene expression vector and a curated vocabulary of $P$ gene programs (the number of curated gene sets in the vocabulary; $220$ for ABA, $P = 318$ for KEGG, $170$ for Reactome), from which it computes a precomputed scalar activation score per program (Eq.~\ref{eq:score}). It outputs (i)~a $P \times P$ attention matrix encoding the pairwise directed dependency from each program to every other program and (ii)~a cell-type label. Stage~1 learns the attention matrix via self-supervised masked reconstruction combined with the intervention-consistent objective, using no labels. Stage~2 fine-tunes a small classification head on cell-type labels to enable condition-stratified analysis. 

\subsection{Input Representation}

Let $\mathbf{g}_i \in \mathbb{R}_+^G$ denote the library-size-normalized, log1p-transformed expression vector of nucleus $i$ over $G$ genes. Let $\Pi = \{\pi^1, \ldots, \pi^P\}$ be a vocabulary of $P$ gene programs, each $\pi_p \subseteq \{1, \ldots, G\}$ a curated gene set. The size-normalized activation score for program $p$ in nucleus $i$ is:
\begin{equation}
  s_{i,p} = \frac{1}{\sqrt{|\pi_p|}} \sum_{g \in \pi_p} g_{i,g},
  \label{eq:score}
\end{equation}
yielding $\mathbf{s}_i \in \mathbb{R}_+^P$. The square-root denominator corrects for program size without penalizing broadly expressed programs.

\subsection{Model Architecture}
\begin{figure}[t]
\centering
\includegraphics[width=\linewidth]{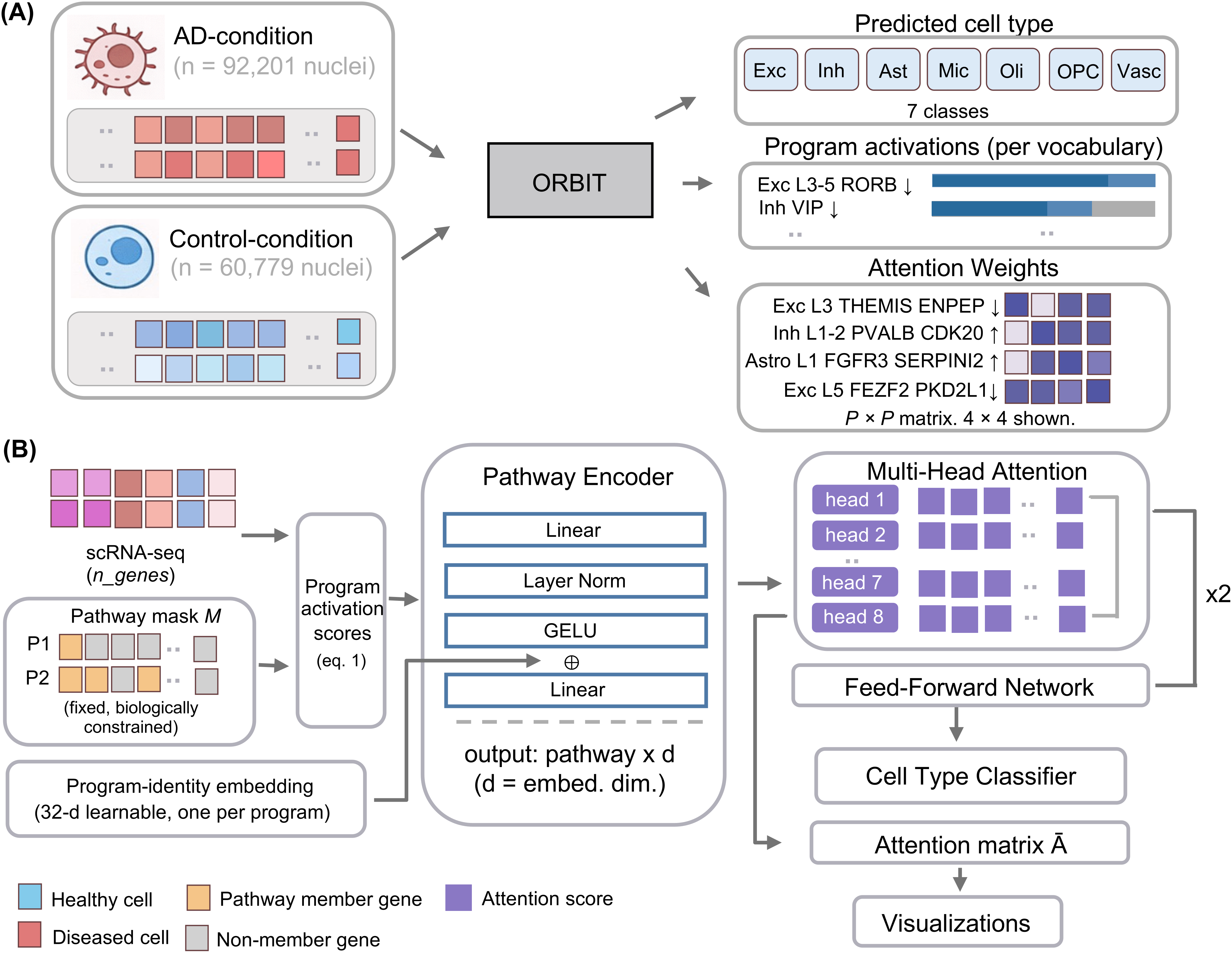}
\caption{
\textbf{ORBIT architecture and information flow.} ORBIT is a two-stage self-supervised model that learns biological program co-activation structure from scRNA-seq data. \textbf{a}, Model overview. ORBIT is trained on 191{,}890 prefrontal cortex nuclei from the Morabito 2021 atlas, comprising AD-condition donors ($n = 92{,}201$) and control-condition donors ($n = 60{,}779$); an 80/20 stratified split (seed 42) reserves 38{,}910 nuclei for held-out evaluation. \textbf{b}, Internal architecture. Raw gene expression is reduced to $P$ scalar program activation scores via the binary pathway membership mask $M$. The Pathway Encoder projects each scalar score into a $D$-dimensional token embedding (with a learnable program-identity component). The resulting sequence is processed by two stacked pre-norm transformer blocks ($\times 2$; 8-head self-attention followed by a feed-forward network). The head-averaged attention matrix $\tilde{A}$ from the final block is averaged across nuclei to yield $\bar{A}$, the attention matrix; the mean-pooled transformer output yields the cell embedding, from which a classification head produces cell-type predictions across 7 classes.}
\label{fig:architecture}
\end{figure}

The architecture has three components: a pathway encoder, a pathway attention transformer, and a classification head (Fig.~\ref{fig:architecture}).

\textbf{Pathway encoder.}
Each scalar score $s_p$ is projected to a 96-dimensional vector via a linear layer, layer normalization, and GELU activation (weights shared across programs). A 32-dimensional learnable embedding assigns each program a unique identity, serving as positional encoding.
The two vectors are concatenated and projected to $D = 128$ dimensions with layer normalization, GELU, and dropout ($p = 0.30$), yielding the token matrix $H^0 \in \mathbb{R}^{P \times D}$.

\textbf{Pathway attention transformer.}
$H^0$ is processed by two stacked pre-norm transformer blocks:
\begin{align}
  H' &= H + \mathrm{MHA}(\mathrm{LayerNorm}(H)), \\
  H'' &= H' + \mathrm{FFN}(\mathrm{LayerNorm}(H')),
\end{align}
where MHA (Multi-Head Attention) denotes 8-head self-attention (attention dropout $p = 0.10$) and FFN (Feed-Forward Network) is a two-layer network with expansion factor four, GELU, and dropout ($p = 0.10$).
The output is layer-normalized and mean-pooled over programs to yield a cell embedding $e_i \in \mathbb{R}^D$.
 
The attention weight tensor $A \in \mathbb{R}^{B \times H \times P \times P}$ of the final block is the primary scientific output.
Averaged over nuclei and heads:
\begin{equation}
  \bar{A}_{p,q} = \frac{1}{NH} \sum_{i=1}^N \sum_{h=1}^H A^{(i,h)}_{p,q},
  \label{eq:Abar}
\end{equation}
where $\bar{A}_{p,q}$ quantifies the mean directional dependency from program $p$ toward program $q$. We distinguish three quantities used throughout: $A^{(i,h)} \in \mathbb{R}^{P \times P}$ denotes the raw per-nucleus, per-head attention map; $\tilde{A}^{(i)} = \frac{1}{H}\sum_h A^{(i,h)}$ denotes its head-average (used in $\mathcal{L}_\text{interv}$ with gradient flow to the final block); and $\bar{A}$ denotes the dataset-level mean used as the scientific output (Eq.~\ref{eq:Abar}).

\textbf{Classification head.}
In Stage~2, a single linear layer mapping $e_i \in \mathbb{R}^D$ to $n_\text{cls} = 7$ logits (where $n_\text{cls}$ is the number of cell-type classes; $16{,}000$ parameters; $3.9\%$ of Stage~1) is appended. Its small capacity prevents label signal from distorting Stage~1 co-activation structure.

\subsection{Stage~1: Self-Supervised Co-Activation Learning}
 
Stage~1 combines three objectives. We describe each in turn.
 
\textbf{Masked reconstruction.}
At each step, 40\% of program scores per nucleus are randomly zeroed.
A linear decoder reconstructs the masked scores:
\begin{equation}
  \mathcal{L}_\text{recon} = \frac{1}{|M|} \sum_{(i,p) \in M} (\hat{s}_{i,p} - s_{i,p})^2.
\end{equation}
This teaches the transformer to infer missing programs from their neighbors, serving as the basic co-activation signal.
 
\textbf{Entropy regularization.}
To prevent attention from collapsing to a single program or dispersing uniformly:
\begin{equation}
  \mathcal{L}_\text{entropy} = \mathbb{E}_p \!\left[ \left( \frac{H(A_p)}{\log P} - 0.5 \right)^{\!2} \right],
\end{equation}
where $H(A_p) = -\sum_q a_{p,q} \log(a_{p,q} + \varepsilon)$ is the row entropy.
 
\textbf{Intervention-consistent influence loss.}
This is the core mechanism that moves ORBIT beyond symmetric co-expression. It operates entirely in \emph{program score space}, forcing the attention weight $A[p,q]$ to predict how program~$q$'s reconstructed score changes when program~$p$'s input is removed.

At each training step, $n_t = 4$ programs are sampled uniformly as
intervention targets.  For each target~$p$, two forward passes produce reconstructed score vectors:
\begin{enumerate}[leftmargin=*, itemsep=1pt]
  \item A \emph{full pass} yielding scores
    $\hat{\mathbf{s}} = d\bigl(f(\mathbf{s})\bigr)$.
  \item An \emph{intervened pass} with $s_p$ set to zero, yielding
    $\hat{\mathbf{s}}^{\setminus p}$.
\end{enumerate}
The observed score-space effect is
$\Delta s_q^{(p)} = \hat{s}_q - \hat{s}_q^{\setminus p}$ (treated as a
stop-gradient target), and the predicted effect is simply:
\begin{equation}
  \Delta \hat{s}_q^{(p)} = A[p,q] \cdot s_p\,,
\end{equation}
where $A[p,q] := \tilde{A}^{(i)}_{p,q}$ is the head-averaged attention weight (final block) from the intervened forward pass, recomputed with gradient so that the query/key parameters of that block receive direct supervision.
The loss is a variance-normalized mean squared error (MSE):
\begin{equation}
  \mathcal{L}_\text{interv}
  = \frac{1}{n_t}\sum_{p}
    \mathrm{MSE}\!\left(
      \frac{\Delta\hat{\mathbf{s}}^{(p)}}{\sigma_{\Delta}},\;
      \frac{\Delta\mathbf{s}^{(p)}}{\sigma_{\Delta}}
    \right),
\end{equation}
where $\sigma_{\Delta}$ is the standard deviation of the observed delta across the batch (clamped for numerical stability).

\textbf{Full Stage~1 objective.}
\begin{equation}
  \mathcal{L}_\text{Stage1} = \mathcal{L}_\text{recon}
  + 0.02\,\mathcal{L}_\text{entropy}
  + 0.05\,\mathcal{L}_\text{interv}.
\end{equation}
Both auxiliary losses are kept an order of magnitude below $\mathcal{L}_\text{recon}$.

\subsection{Stage~2: Cell-Type Classification}
 
Stage~2 proceeds in two phases. During phase~1 (linear probe), Stage~1 weights are frozen; only the classification head is updated for up to 50 epochs with early stopping (patience 5). In phase~2 (end-to-end), all weights are unfrozen for 20 additional epochs. The gain from Phase~1 to Phase~2 quantifies how much label supervision improves over the self-supervised representation.

\subsection{Rewiring Analysis and Gene-Level Projection}

After training, rewiring analysis is purely post-hoc and involves no additional learning.
 
\textbf{Rewiring map.}
Per cell type $ct$, the signed delta $\Delta\bar{A}_{ct} = \bar{A}_{ct}^{\mathrm{AD}} - \bar{A}_{ct}^{\mathrm{ctrl}}$ identifies gained and lost co-activation.
Statistical significance is assessed by permuting condition labels 1{,}000 times with Benjamini--Hochberg FDR correction per directed pair.
All reported pairs with $|\Delta\bar{A}| \ge 0.004$ survive FDR at $q < 0.05$ across all three vocabularies.
 
\textbf{Gene-level projection.}
To connect program-level rewiring to individual genes, we define a gene-pathway influence matrix:
\begin{equation}
  I_{g, p_2} = \sum_{p_1} M_{g,p_1} \cdot \bar{A}_{p_1, p_2},
  \label{eq:influence}
\end{equation}
where $M \in \{0,1\}^{G \times P}$ is the binary gene-program membership matrix.
Substituting $\Delta\bar{A}$ yields $\Delta\mathbf{I}$, identifying genes most implicated in rewiring.

\section{Experiments}

\subsection{Data and Implementation}

\textbf{Data.}
Stage~1 uses the Morabito 2021 snRNA-seq atlas (GSE174367; 191{,}890 nuclei, seven cell types, prefrontal cortex)~\citep{morabito2021}. Cross-dataset replication uses a held-out disease cohort~\citep{lau2020} and a control dataset (MAP2$^+$ neuron-enriched, 20{,}000 nuclei)~\citep{otero2022}. Three pathway vocabularies are evaluated: Allen Brain Atlas (ABA)~\citep{shen2012} (220 programs), Kyoto Encyclopedia of Genes and Genomes (KEGG)~\citep{kanehisa2000} (318 programs), and Reactome~\citep{joshitope2005} (170 programs), each filtered to $\geq\!5$ member genes. A separate ORBIT model is trained per vocabulary; findings replicated across all three are reported as primary results.

\textbf{Training.} ORBIT is implemented in PyTorch~2.1. Stage~1 uses AdamW ($\beta_1 = 0.9$, $\beta_2 = 0.999$, weight decay $= 0.01$), cosine schedule with 10\% linear warmup (peak lr $= 10^{-4}$), gradient clipping at norm~$1.0$, batch size 128, 50 epochs. Stage~2 Phase~2 uses lr~$= 5 \times 10^{-5}$ with cosine warmup; inverse-frequency class weighting and oversampling address class imbalance (vascular cells $<1\%$). Total parameters: Stage~1 $\approx 415{,}000$; Stage~2 head $\approx 16{,}000$.

\subsection{Synthetic Validation}
\label{subsection:synthetic}
\begin{table}[t]
\centering
\caption{%
  \textbf{Hidden triplet recovery on synthetic data.}
  Four AND-gate interactions $(A,B){\to}C$ embedded among 40 programs with $|\rho(A,C)|{=}|\rho(B,C)|{=}0$ by construction.
  \emph{Dir.\ Acc.}: fraction of true edges correctly oriented;
  \emph{ASI}: mean signed asymmetry index (positive $=$ correct);
  \emph{Recovery}: fraction of triplets with both parents in top 10\%.
  Shaded: symmetric methods.
}
\label{tab:synthetic}
\smallskip
\begin{tabular}{@{}l ccc@{}}
\toprule
Method & Dir.\ Acc. & ASI & Recovery \\
\midrule
\textbf{ORBIT (with $\mathcal{L}_\text{interv}$)}
  & \textbf{1.00} & $\mathbf{+0.78}$ & \textbf{0.75} \\
\quad without $\mathcal{L}_\text{interv}$
  & 0.00 & $-0.57$ & 0.00 \\
\addlinespace[2pt]
Random Forest
  & 0.88 & $+0.38$ & \textbf{0.75} \\
MLP
  & 0.75 & $+0.32$ & 0.50 \\
\addlinespace[2pt]
\rowcolor[gray]{0.95}
Pearson $|\rho|$ & 0.00 & 0.00 & 0.00 \\
\rowcolor[gray]{0.95}
Mutual Info     & 0.00 & 0.00 & 0.00 \\
\rowcolor[gray]{0.95}
Distance Corr   & 0.00 & 0.00 & 0.00 \\
\rowcolor[gray]{0.95}
HSIC            & 0.00 & 0.00 & 0.00 \\
\bottomrule
\end{tabular}
\end{table}

Before applying ORBIT to biological data, we verify in a controlled setting that the intervention-consistent objective is both necessary and sufficient for recovering directed, non-additive program dependencies invisible to correlation.
We construct a synthetic dataset ($N\!=\!8{,}000$ cells, $G\!=\!2{,}000$ genes, $P\!=\!40$ programs) containing 12 additive co-activation edges (recoverable by Pearson) and 4 multiplicative AND-gate $(A, B, C)$, in which the activation of program $C$ is conditionally modulated only when both $A$ and $B$ exceed their respective median activity levels.
The interaction is orthogonalized against linear main effects (Appendix~\ref{app:synthetic_design}), guaranteeing that the dependency is genuinely non-additive and invisible to any pairwise statistic.
 
ORBIT with $\mathcal{L}_\text{interv}$ recovers 75\% of triplets with perfect directionality (Dir.\ Acc.${=}1.00$, ASI${=}{+}0.78$; Table~\ref{tab:synthetic}). All four symmetric baselines recover 0\%. Removing $\mathcal{L}_\text{interv}$ collapses both recovery and asymmetry and \emph{reverses} the directional signal (ASI${=}{-}0.57$), despite identical architecture and reconstruction performance, confirming that the intervention objective is what induces correctly oriented directional structure.

\subsection{Cell-Type Classification}

As a secondary validation, we assess whether the learned co-activation representation preserves cell-type identity. Unless otherwise stated, classification performance is reported as macro F1. On an 80/20 stratified split (seed 42), ORBIT achieves macro F1$\,{=}\,0.984$ from $P{=}220$ pathway scores (Table~\ref{tab:classification}).
CellTypist~\citep{dominguez2022}, operating on all 22{,}088 genes under the same split, achieves $0.987$, which is a 0.3-point gap despite ORBIT's $100{\times}$ reduction in input dimensionality. To isolate the contribution of the attention mechanism, we train an MLP on the same pathway scores, which reaches $0.979$. The additional 0.5-point gain from ORBIT indicates that its architecture extracts predictive structure from the pathway representation beyond what a feedforward model recovers. These results show that the co-activation structure retains sufficient information for downstream classification, despite substantial compression to pathway-level inputs.

\begin{table}[t]
\centering
\caption{Cell-type classification on Morabito 2021 (80/20 stratified split, seed 42). 
This table reports supervised models operating under the same evaluation protocol. Macro F1 is averaged over six major cell types; vascular cells are oversampled and excluded from macro averaging (see Methods).}
\label{tab:classification}
\small
\begin{tabular}{lcccc}
\toprule
Method & Macro F1 & Exc F1 & Inh F1 & Mic F1 \\
\midrule
\textbf{ORBIT (this work)} & \textbf{0.984} & \textbf{0.977} & \textbf{0.985} & \textbf{0.976} \\
CellTypist~\citep{dominguez2022} & 0.987 & 0.979 & 0.988 & 0.989 \\
scANVI~\citep{xu2021} & 0.990 & 0.990 & 1.000 & 1.000 \\
MLP Baseline & 0.979 & 0.980 & 0.984 & 0.986 \\
\bottomrule
\end{tabular}
\end{table}

\subsection{Co-Activation Structure in Healthy Tissue}
\begin{figure}[t]
\centering
\includegraphics[width=\linewidth]{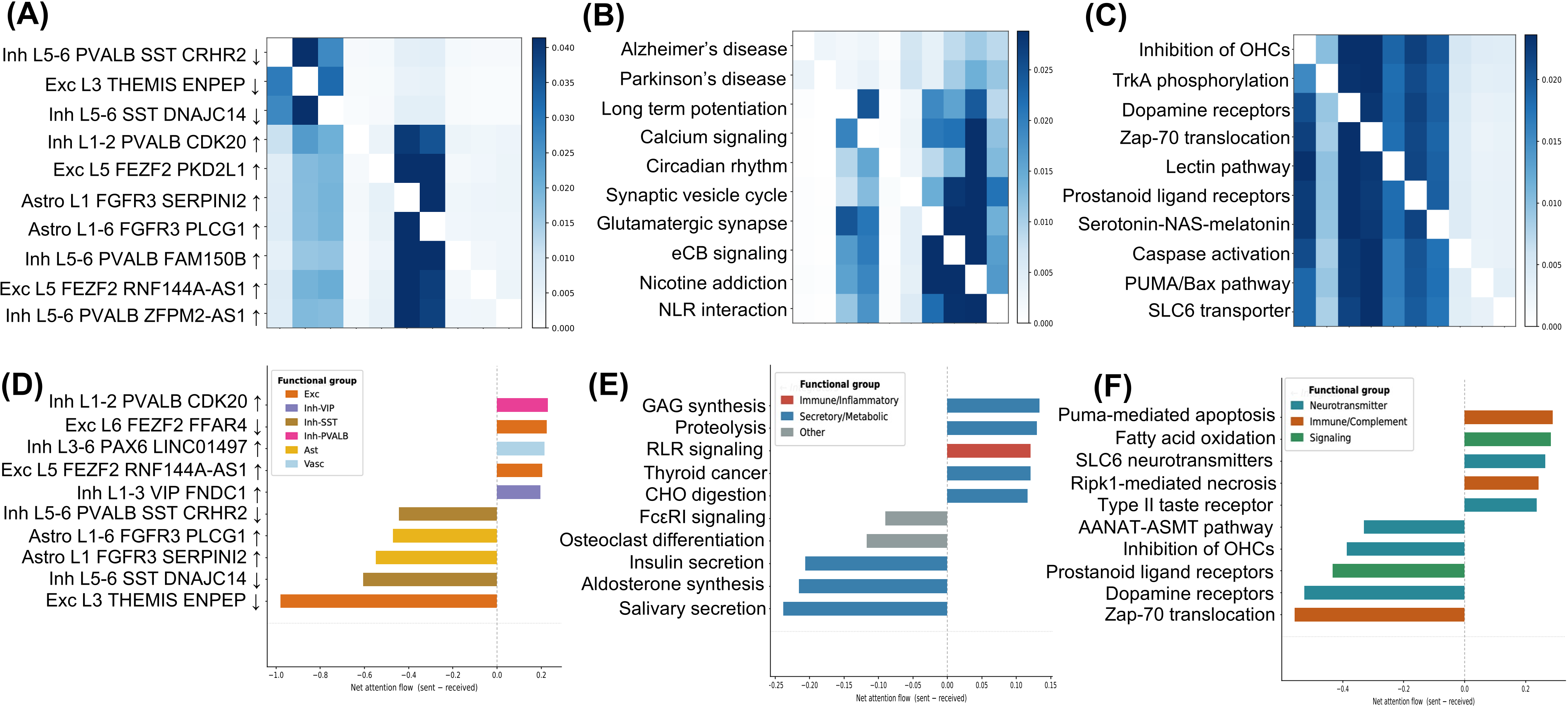}
\caption{
\textbf{ORBIT integrates multi-dataset cortical transcriptomes and learns vocabulary-robust inter-program attention structure.} \textbf{a}, Mean inter-program attention heatmap for the top 10 highest-variance ABA programs. \textbf{b}, KEGG vocabulary replication: mean attention heatmap. \textbf{c}, Reactome vocabulary replication: mean attention heatmap. Color scales in (a–c) represent mean attention weight. \textbf{d}, Directed attention: transcriptional program drivers vs.\ integrators of ABA programs. \textbf{e}, KEGG vocabulary replication: directed attention. \textbf{f}, Reactome vocabulary replication: directed attention.
}
\label{fig:healthy_heatmaps}
\end{figure}

\textbf{Cross-vocabulary consistency.}
Across all three vocabularies, ORBIT recovers cell-type-specific co-activation patterns consistent with established biology (Fig.~\ref{fig:healthy_heatmaps}a--f). In the ABA vocabulary, PVALB and FEZF2 excitatory upregulated neuronal subtypes (Exc L5 FEZF2 PKD2L1, Exc L5 FEZF2 RNF144A-AS1, Inh PVALB CDK20) are net attention drivers, while excitatory and interneuron downregulated programs (Exc L3 THEMIS ENPEP, Inh L5-6 SST DNAJC14) and astrocyte programs are net integrators (Fig.~\ref{fig:healthy_heatmaps}a,d), which is consistent with cortical feedforward inhibition in which pyramidal neurons recruit local interneurons~\citep{buzsaki1984}. In KEGG, Long-term Potentiation and Calcium Signaling form a tightly coupled pair~\citep{connor1999calcium}, and Glycosaminoglycan Biosynthesis acts as a net driver (Fig.~\ref{fig:healthy_heatmaps}b,e), consistent with its role in extracellular matrix regulation of synaptic plasticity~\citep{ali2025glycosaminoglycans}. In the Reactome vocabulary, presynaptic and neurotransmitter programs (Serotonin and Melatonin Biosynthesis (abbreviated as "AANAT-ASMT pathway")~\citep{pandiperumal2008physiological}; Dopamine Receptors~\citep{pan2019dopamine} and Na$^+$/Cl$^-$-dependent Neurotransmitter Transporters (abbreviated as "SLC6 neurotransmitters")~\citep{ayka2020slc}) form the highest-attention cluster (Fig.~\ref{fig:healthy_heatmaps}c,f), consistent with the enrichment of synaptic machinery in cortical neurons~\citep{dewilde2016meta}.

\textbf{Gene-level influence.}
Top-ranked genes of the gene-pathway influence matrix reveal what biological features the attention graph has learned to prioritize. Interestingly, in ABA, top-ranked influence genes are dominated by synaptic organizers and neuronal adhesion molecules (\textit{LRRTM4}, \textit{DAB1}, \textit{FLRT2}, \textit{CDH10}, and \textit{NRXN2}), routing primarily into excitatory and interneuron integrator programs (Exc L3 THEMIS ENPEP, Inh L5-6 SST DNAJC14)~\citep{agosto2021,farini2021,fleitas2021,laszlo_2022,lin2023}. A second cluster led by \textit{GSAP} ($\gamma$-secretase activating protein, which directly promotes APP cleavage~\citep{jin2022}) and \textit{NRXN2} shows highest influence on astrocyte FGFR3 programs (SERPINI2, AQP1, PLCG1), unsupervised recovery of a known AD-relevant gene routing into a glial program~\citep{medina2023}.

In KEGG, where vocabulary scope is organized around signaling cascades rather than cell identity, influence rankings are instead dominated by hub kinases (\textit{MAPK1}, \textit{AKT1/3}, \textit{PIK3CA})~\citep{dinsmore2018mapk}; in Reactome, by proteostasis machinery (\textit{TP53}, \textit{PSMD1/3})~\citep{aubrey2017p53,bencomo2021psmd}.
These are not canonical cell-type markers, but they are biologically coherent with each vocabulary's domain. These results confirm that each vocabulary's influence landscape is internally coherent: the attention graph adapts to the biological scope of its input gene sets (Appendix Fig.~\ref{fig:gene_to_pathway}).

\subsection{AD Rewiring of Co-Activation Structure}
\begin{figure}[t]
\centering
\includegraphics[width=\linewidth]{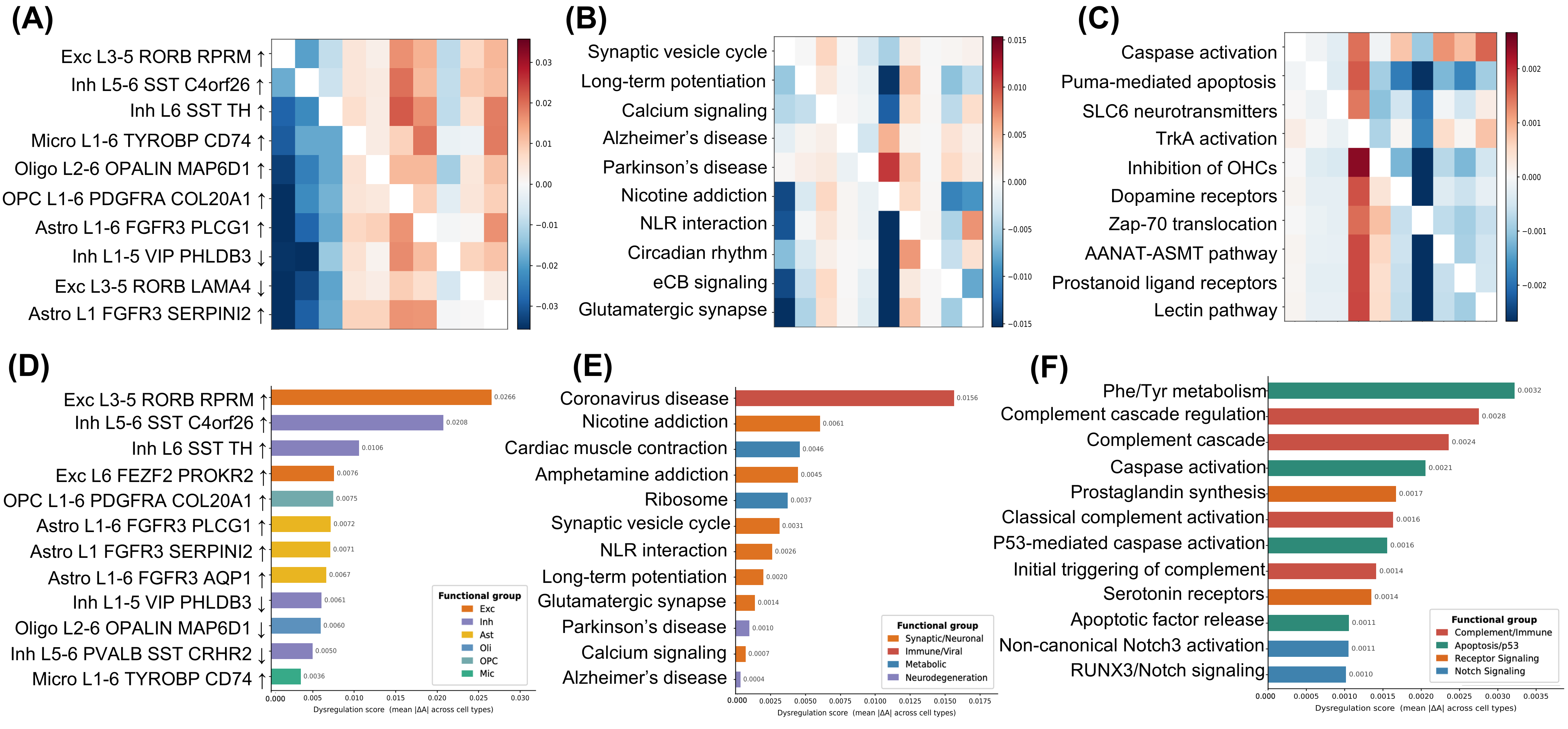}
\caption{
\textbf{AD rewires inter-program attention in a cell-type-specific and gene-anchored manner.} \textbf{a}, Attention heatmap for the top 10 programs by absolute $\Delta$attention (ABA). \textbf{b}, Top 10 programs by $\Delta$attention (KEGG). \textbf{c}, Top 10 programs by $\Delta$attention (Reactome). Color scales in (a–c) represent absolute Δattention. \textbf{d}, Top 10 dysregulated programs by summed outgoing $|\Delta\text{attention}|$ (ABA). \textbf{e}, Cell-type vulnerability ranking (mean $|\Delta\text{attention}|$ per cell type). \textbf{f}, Pathway co-activation rewiring: gained or lost connection in AD (scatter plots by cell type).
}

\label{fig:rewiring}
\end{figure}

Comparing $\bar{A}_{\mathrm{AD}}$ vs.\ $\bar{A}_{\mathrm{ctrl}}$ within each cell type reveals two consistent rewiring clusters (Figs.~\ref{fig:rewiring}a--f, \ref{fig:gene_to_pathway}). 

\textbf{Gained co-activation.} In ABA, RORB$^+$ excitatory neurons spanning cortical layers 3-5 (the "Exc L3-5 RORB RPRM" pathway in ABA nomenclature) display the strongest overall co-activation restructuring in AD, with total attention change of 0.0296 summed across all target programs. This enrichment is consistent with the fact that the deep-layer RORB$^+$ population is previously identified as selectively vulnerable in AD~\citep{leng2021}. In KEGG, Alzheimer disease and Parkinson disease programs emerge as a mutual-attention pair, consistent with prior studies that identified overlapping risk loci between Alzheimer’s disease and Parkinson’s disease~\citep{wainberg2023shared}. In Reactome, the TrkA activation program shows a divergent rewiring pattern: its outgoing attention into early-target columns increases in AD, while other source programs in the same set lose attention to those targets. TrkA is the high-affinity NGF receptor, and NGF–TrkA signaling is impaired in Alzheimer’s disease~\citep{mufson2008cholinergic}.
 
\textbf{Lost co-activation.} The most striking finding across vocabularies is selective loss of astrocyte--interneuron co-activation (ABA: Astro FGFR3 $\to$ Inh SST/TH pairs; maximum $|\Delta\bar{A}| = 0.03$).
This loss is confined to program pairs with the \emph{highest baseline control weights}, suggesting that the most strongly coupled healthy-state pairs decouple preferentially. Gene-level projection confirms highest negative dysregulation in SST and PVALB marker genes, consistent with established interneuron vulnerability~\citep{xu2020,habib2020}.
 
\textbf{Cell-type vulnerability.}
Excitatory neurons rank highest by mean $|\Delta\bar{A}|$, followed by inhibitory interneurons and astrocytes, consistent across all three vocabularies and with the clinical primacy of excitatory neuron loss in AD cortex~\citep{buzsaki1984}.

\subsection{Pathway Score Validation}
\textbf{Agreement with AUCell.}
ORBIT pathway scores achieve median Pearson $r = 0.922$ against AUCell-derived rAUC scores~\citep{aibar2017} across 220 ABA programs, with 52.3\% exceeding $r > 0.9$. Agreement is strong for programs with ${\leq}100$ genes (mean $r = 0.72$--$0.87$) and degrades for larger programs ($r \approx 0.28$), reflecting AUCell's known score dilution for large gene sets. This confirms that the pathway encoder captures genuine program activity.

\textbf{Cross-dataset replication.}
Applied without retraining to a held-out neuron-enriched control dataset, ORBIT achieves Spearman $r_s = 0.642$ between attention matrices from the reference and held-out cohorts. The moderate correlation confirms that dominant co-activation relationships are a property of cortical transcriptional organization, not a single-dataset artifact.

\textbf{Shuffled-label control.}
Permuted cell-type labels reduce Stage~2 accuracy from 98.8\% to 39.8\% (chance = 14.3\%), while the attention matrix retains $r = 0.911$ with the correctly-labeled model, which is expected because Stage~1 is label-free and both models converge on the same expression covariance structure.

\section{Discussion}
\label{sec:discussion}

\textbf{Biological interpretation of ORBIT.} Three findings support the interpretation of $\bar{A}$ as a biologically meaningful, cell-type-stratified summary of program co-activation. First, the same trained model produces qualitatively distinct attention matrices across cell types (Fig.~\ref{fig:allen_pathway_activation}--\ref{fig:reactome_pathway_activation}). Since Stage~1 receives no cell-type labels, the cell-type structure is recovered from expression covariance alone. Second, the gene-pathway influence map (Eq.~\ref{eq:influence}, Fig.~\ref{fig:gene_to_pathway}) ranks biologically coherent genes within each vocabulary's domain without any external prior on which genes should be central. Third, the disease-versus-control rewiring map (Section~\ref{sec:method}, Fig.~\ref{fig:rewiring}) recovers patterns previously reported in the AD literature. The candidate co-activation relationships ORBIT recovers are hypotheses, not validated mechanisms. Their value is in narrowing the search space for follow-up experiments, including CRISPR perturbation, co-immunoprecipitation, or spatial co-localization, to a small number of directed pairs with prior biological plausibility.

\textbf{Scope of directionality and attention-based interpretation.}
A standard concern with attention-based interpretation is that attention weights are not trained to be explanations: they emerge as a byproduct of some other task and are inspected after the fact~\citep{jainwallace2019attention,wiegreffepinter2019attention}. ORBIT inverts this. The intervention-consistent objective trains attention weights directly: for each program pair $(A, B)$, the model must use the weight from $A$ to $B$ to predict how $B$'s score shifts when $A$ is removed from the input. The prediction is simply the attention weight times the input score, with no learnable parameters between them, so the weight has to track the observed shift, otherwise the loss is nonzero. This converts attention from a post-hoc summary into a trained quantity with a defined predictive role, and it produces a structural property no symmetric measure can: the forward weight from $A$ to $B$ and the backward weight from $B$ to $A$ are genuinely distinct, where Pearson correlation, partial correlation, factor decomposition, and every other symmetric statistic produce this distinction as identically zero. The synthetic experiments in Section~\ref{subsection:synthetic} confirm this is not a vacuous distinction: ORBIT recovers the correct directionality on hidden non-additive triplets where every symmetric baseline scores zero.

\textbf{Limitations and future directions.}
ORBIT has three limitations worth noting. First, the model aggregates attention over nuclei within each condition and cell type, which stabilizes the $P \times P$ estimate but precludes modeling nucleus-to-nucleus variability in co-activation. Disease-associated heterogeneity within an annotated cell type is therefore not directly observable, and a hierarchical extension learning per-nucleus attention distributions alongside population means is a natural next step. Second, effect sizes are also small in absolute units ($|\Delta\bar{A}| \leq 0.03$), but this magnitude is appropriate to the scale of the problem: for instance, with $P = 220$ programs in the ABA vocabulary, uniform attention assigns approximately $1/220 \approx 0.0045$ to each target, so a $\Delta\bar{A}$ of 0.03 corresponds to a substantial reorganization of attention weights. Additionally, all reported pairs survive three independent safeguards (permutation FDR $q < 0.05$, Jaccard stability $\geq 0.70$ across subsamples, and replication across vocabularies). Third, generalization beyond prefrontal cortex AD, including to other brain regions, disease contexts, or non-neural tissues, is an open empirical question. ORBIT's framework is vocabulary- and architecture-agnostic, and the intervention-consistent objective applies wherever features can be partitioned into meaningful groups and masked reconstruction is a valid self-supervised objective.

\section{Conclusion}
We introduced ORBIT, a two-stage attention-based framework for learning directed, asymmetric gene program co-activation structure from single-cell RNA-sequencing data. Applied to an Alzheimer's disease snRNA-seq atlas across three pathway vocabularies, ORBIT recovers co-activation patterns consistent with established AD vulnerability signatures and identifies directed program-level rewiring events structurally undetectable by differential expression or graph-constrained methods. ORBIT is vocabulary-agnostic, positioning it as a scalable approach for directed pathway rewiring analysis across disease contexts.



\bibliographystyle{plainnat}
\bibliography{references}

\newpage
\appendix

\section{Training Hyperparameters}
\label{app:hyperparams}

\subsection{Stage~1 Hyperparameters}

\begin{table}[h]
\centering
\caption{Stage~1 self-supervised training hyperparameters (all three 
vocabulary models).}
\label{tab:hparams_s1}
\small
\begin{tabular}{lll}
\toprule
Hyperparameter & Value & Notes \\
\midrule
Optimizer & AdamW & $\beta_1 = 0.9$, $\beta_2 = 0.999$ \\
Weight decay & 0.01 & \\
Peak learning rate & $1 \times 10^{-4}$ & Cosine schedule \\
Warmup & 10\% of epochs & Linear ramp \\
Gradient clipping & norm $= 1.0$ & \\
Batch size & 128 & \\
Epochs & 50 & \\
Masking ratio & 0.40 & 40\% of programs zeroed per nucleus \\
Intervention targets $n_t$ & 4 & Programs sampled per step \\
$\mathcal{L}_\text{entropy}$ weight & 0.02 & \\
$\mathcal{L}_\text{interv}$ weight & 0.05 & \\
$\mathcal{L}_\text{recon}$ weight & 1.00 & \\
\bottomrule
\end{tabular}
\end{table}

\subsection{Stage~2 Hyperparameters}

\begin{table}[h]
\centering
\caption{Stage~2 cell-type classification hyperparameters.}
\label{tab:hparams_s2}
\small
\begin{tabular}{lll}
\toprule
Hyperparameter & Value & Notes \\
\midrule
Phase~1 (linear probe) & up to 50 epochs & Early stopping, patience 5 \\
Phase~2 (end-to-end) & 20 epochs & \\
Learning rate (Phase~2) & $5 \times 10^{-5}$ & Cosine warmup \\
Class weighting & Inverse-frequency & \\
Oversampling & Vascular cells only & ${<}1\%$ of nuclei \\
Random seed & 42 & Train/test split \\
\bottomrule
\end{tabular}
\end{table}

\subsection{Model Architecture Summary}

\begin{table}[h]
\centering
\caption{ORBIT parameter counts by component (ABA vocabulary, $P = 220$).}
\label{tab:params}
\small
\begin{tabular}{lrr}
\toprule
Component & Parameters & \% of Total \\
\midrule
Pathway encoder & 71{,}040 & 17.1\% \\
Transformer block 1 & 132{,}352 & 31.9\% \\
Transformer block 2 & 132{,}352 & 31.9\% \\
Reconstruction decoder & 79{,}360 & 19.1\% \\
\textbf{Stage~1 total} & \textbf{415{,}104} & \textbf{100\%} \\
\midrule
Classification head & 16{,}000 & 3.9\% of Stage~1 \\
\bottomrule
\end{tabular}
\end{table}

\section{Compute Resources}
\label{app:compute}

All ORBIT models were trained on a single NVIDIA A100 40 GB GPU using 
PyTorch~2.1. The three vocabulary models (ABA, KEGG, Reactome) were trained 
independently. Approximate wall-clock times are shown in 
Table~\ref{tab:compute}.

\begin{table}[h]
\centering
\caption{Training time and memory usage per vocabulary model 
(single A100 40 GB GPU, batch size 128).}
\label{tab:compute}
\small
\begin{tabular}{lrrr}
\toprule
Vocabulary & $P$ & Stage~1 time & Stage~2 time \\
\midrule
Allen Brain Atlas (ABA) & 220 & ${\approx}1.2$ h & ${\approx}0.3$ h \\
KEGG & 318 & ${\approx}1.2$ h & ${\approx}0.3$ h \\
Reactome & 170 & ${\approx}1.5$ h & ${\approx}0.5$ h \\
\midrule
\textbf{Total} & --- & ${\approx}3.9$ h & ${\approx}1.1$ h \\
\bottomrule
\end{tabular}
\end{table}

\section{Reproducibility Protocol}
\label{app:reproducibility}

\begin{itemize}
\item All three vocabulary models use identical architecture and training pipelines; only the pathway membership matrix $M$ and $P$ differ.
\item Random seeds are fixed for dataset splitting (seed 42), weight initialization, masking, and intervention target sampling.
\item Hyperparameters (Table~\ref{tab:hparams_s1}--\ref{tab:hparams_s2}) were selected on a held-out 10\% validation split of the Morabito reference atlas, fixed before any evaluation on the held-out 20\% test set. No test-set information influenced hyperparameter selection.
\item Rewiring results reported as primary findings must survive all three criteria: $|\Delta\bar{A}| \geq 0.004$, permutation FDR $q < 0.05$, and Jaccard stability $\geq 0.70$ across 10 random 80/20 subsamples.
\end{itemize}

\section{Synthetic Generator Design and Orthogonalization}
\label{app:synthetic_design}

\subsection{Generator Construction}

The synthetic dataset embeds two interaction types among $P = 40$ programs: 12 additive co-activation pairs detectable by Pearson correlation, and 4 multiplicative AND-gate triplets $(A, B) \to C$ in which the effect of either parent depends on the state of the other. The triplets are the focus of this appendix; their construction removes linear pairwise correlations and substantially reduces marginal pairwise signal, making recovery difficult for methods that do not jointly model $s_A$ and $s_B$.

For each triplet, the raw interaction term is
\begin{equation}
  \delta_C = \alpha \cdot
  \mathbf{1}\!\left[s_A > \mathrm{med}_A \wedge s_B > \mathrm{med}_B\right]
  \cdot \varepsilon, \qquad
  \varepsilon \sim \mathrm{Uniform}(-1,1),
\end{equation}
where $\alpha$ controls interaction strength. This term is orthogonalized against the linear main effects of $s_A$ and $s_B$:
\begin{equation}
  \tilde{\delta}_C = \delta_C
  - \frac{\langle \delta_C, s_A \rangle}{\langle s_A, s_A \rangle} s_A
  - \frac{\langle \delta_C, s_B \rangle}{\langle s_B, s_B \rangle} s_B,
\end{equation}
and added to $s_C$, followed by a second projection removing residual linear dependence:
\begin{equation}
  s_C \leftarrow s_C
  - \frac{\langle s_C, s_A \rangle}{\langle s_A, s_A \rangle} s_A
  - \frac{\langle s_C, s_B \rangle}{\langle s_B, s_B \rangle} s_B.
\end{equation}

\textbf{Scope of marginal-signal removal.}
After projection, $\mathrm{corr}(s_A, s_C) = \mathrm{corr}(s_B, s_C) = 0$ in expectation. This eliminates linear pairwise signal but does not imply independence: the AND-gate induces conditional variance differences in $s_C$ that nonlinear pairwise methods (e.g., mutual information, HSIC, distance correlation) could in principle detect. In practice, however, Table~\ref{tab:synthetic} shows that these estimators recover $0\%$ of triplets at $N = 8{,}000$, suggesting that the remaining nonlinear pairwise signal is below their detection threshold at this sample size. Finite-sample effects introduce small residual correlations ($\mathcal{O}(1/\sqrt{N})$), and the projection step yields a noisy, projected interaction rather than a pure multiplicative function. The generator is therefore best viewed as a controlled benchmark that suppresses marginal signal without claiming information-theoretic impossibility.

\subsection{Joint Intervention Term}
The joint-intervention probe is used only in the synthetic setting, where ground-truth triplets are known. Its role is to expose the super-additive interaction component that is only observable when both parents are perturbed simultaneously.

Under single-program intervention, removing $A$ produces a shift $\Delta s_C^{(A)}$ that is conditioned on the state of $B$, since the gate fires only when $s_B$ is high. As a result, single-program signals provide a sparse and condition-dependent view of the interaction. The joint intervention instead directly measures the super-additive residual
\begin{equation}
  \Delta s_C^{(A,B)} - \Delta s_C^{(A)} - \Delta s_C^{(B)},
\end{equation}
on the subset of cells where both parents are active, and uses it as a supervisory signal for both $A[A,C]$ and $A[B,C]$.

For each triplet, four forward passes (full, $-A$, $-B$, $-(A,B)$) yield the corresponding score shifts. The predicted joint effect is
\begin{equation}
  \Delta \hat{s}_C^{(A,B)}
  = A[A,C]\cdot\Delta s_C^{(A)}
  + A[B,C]\cdot\Delta s_C^{(B)},
\end{equation}
and the loss penalizes mismatch with the observed joint shift. When the true interaction is super-additive, the residual is positive and its gradient increases both attention weights beyond what single-program interventions alone would support. In this sense, the joint term directly targets the interaction residual that is difficult to recover from single-program interventions without explicit conditioning or combinatorial stratification.

\textbf{Why the joint term is omitted on biological data.}
Applying the joint term to biological data would require evaluating a large number of candidate triplets ($\binom{P}{2} \cdot P \approx 5.3$M for $P=220$), which is both computationally expensive and statistically underpowered without strong prior filtering. Such filtering would reintroduce predefined interaction structure, contrary to ORBIT’s design goal.

More importantly, the synthetic generator is adversarially constructed to remove marginal signal, whereas real biological data typically retain residual co-activation that single-program interventions can exploit. The joint term is therefore most useful in settings where marginal signal has been deliberately suppressed. We do not claim that single-program intervention fully recovers all higher-order interactions in biological systems; rather, it captures the component that is observable through individual perturbations. The omission of the joint term is thus a pragmatic choice reflecting computational and epistemic constraints, rather than a theoretical guarantee of completeness.

\section{Extended Vocabulary-Specific Results}
Full per-cell-type attention heatmaps for Allen Brain Atlas, KEGG, and Reactome vocabularies; gene influence propagation graphs for each vocabulary; per-class classification metrics; and AUCell correlation stratified by program size bin are provided as supplementary figures and tables.

\subsection{Per-Cell-Type Attention Heatmaps}
\begin{figure}[t]
\centering
\includegraphics[width=\linewidth]{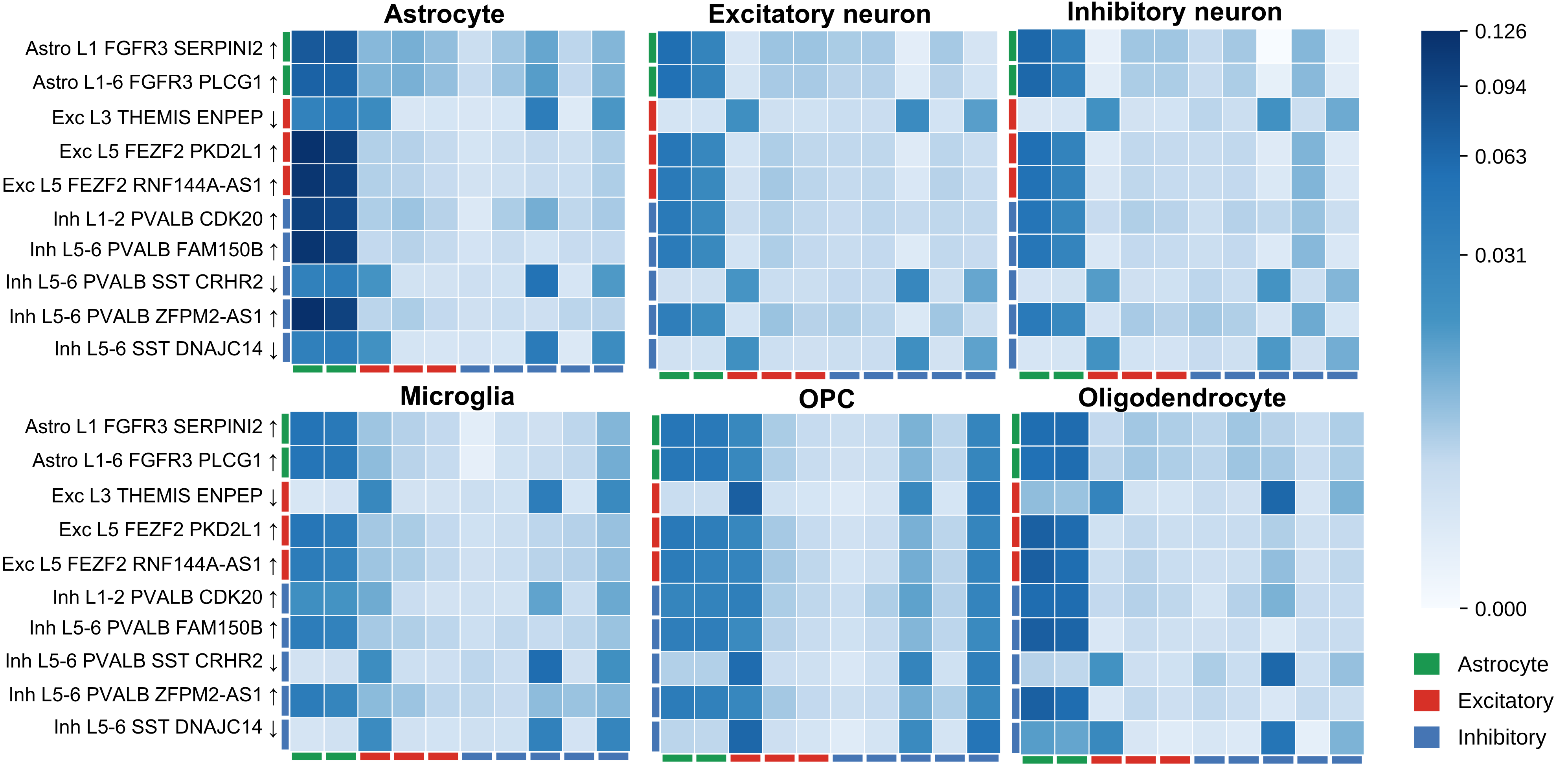}
\caption{
\textbf{Per-cell-type attention heatmaps, Allen Brain Atlas vocabulary.} Mean inter-program attention for the top-variance ABA programs, stratified by cell type.}
\label{fig:allen_pathway_activation}
\end{figure}

\begin{figure}[t]
\centering
\includegraphics[width=\linewidth]{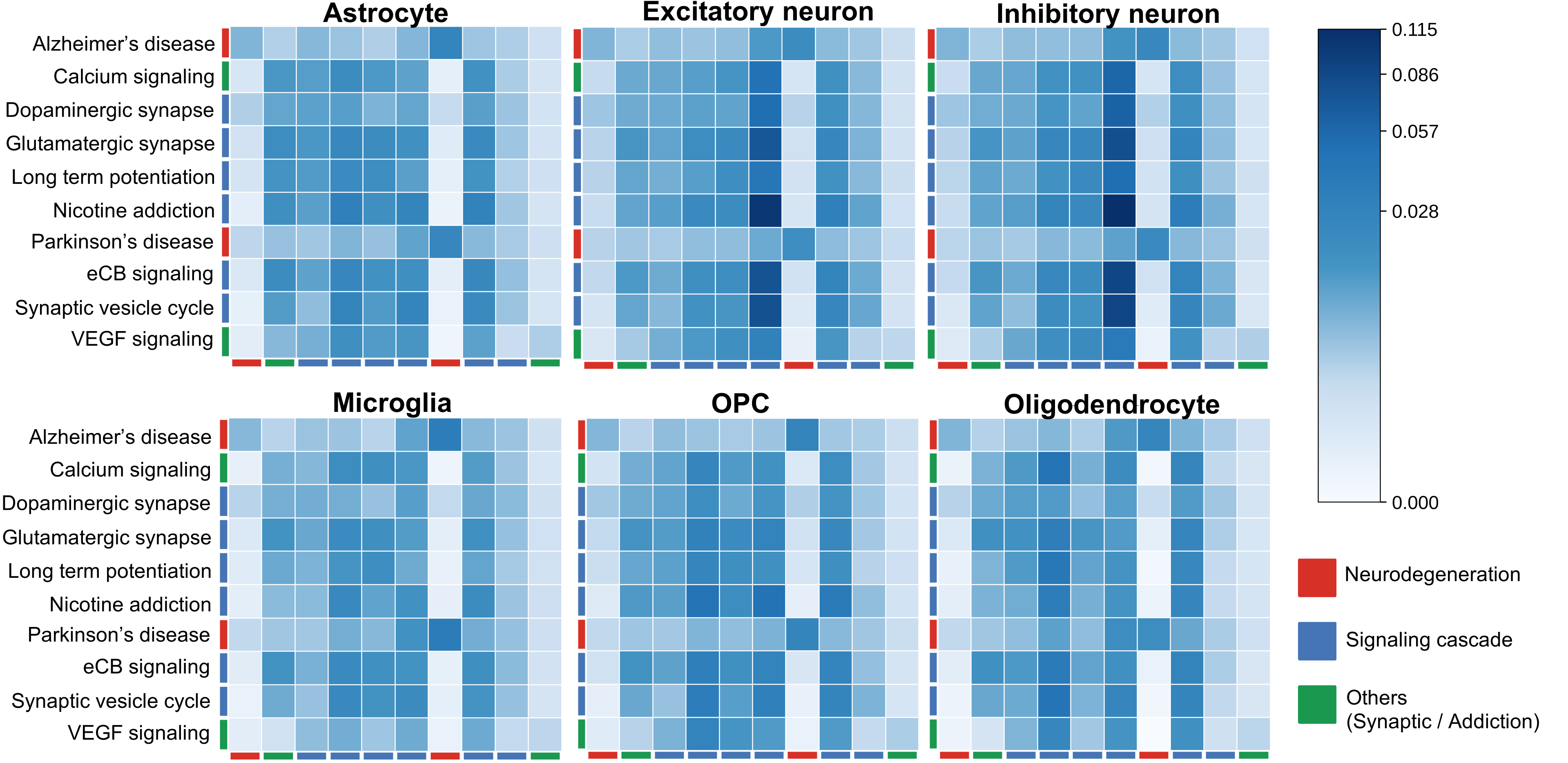}
\caption{
\textbf{Per-cell-type attention heatmaps, KEGG vocabulary. Mean inter-program attention for the top-variance KEGG programs, stratified by cell type.}
}
\label{fig:kegg_pathway_activation}
\end{figure}

\begin{figure}[t]
\centering
\includegraphics[width=\linewidth]{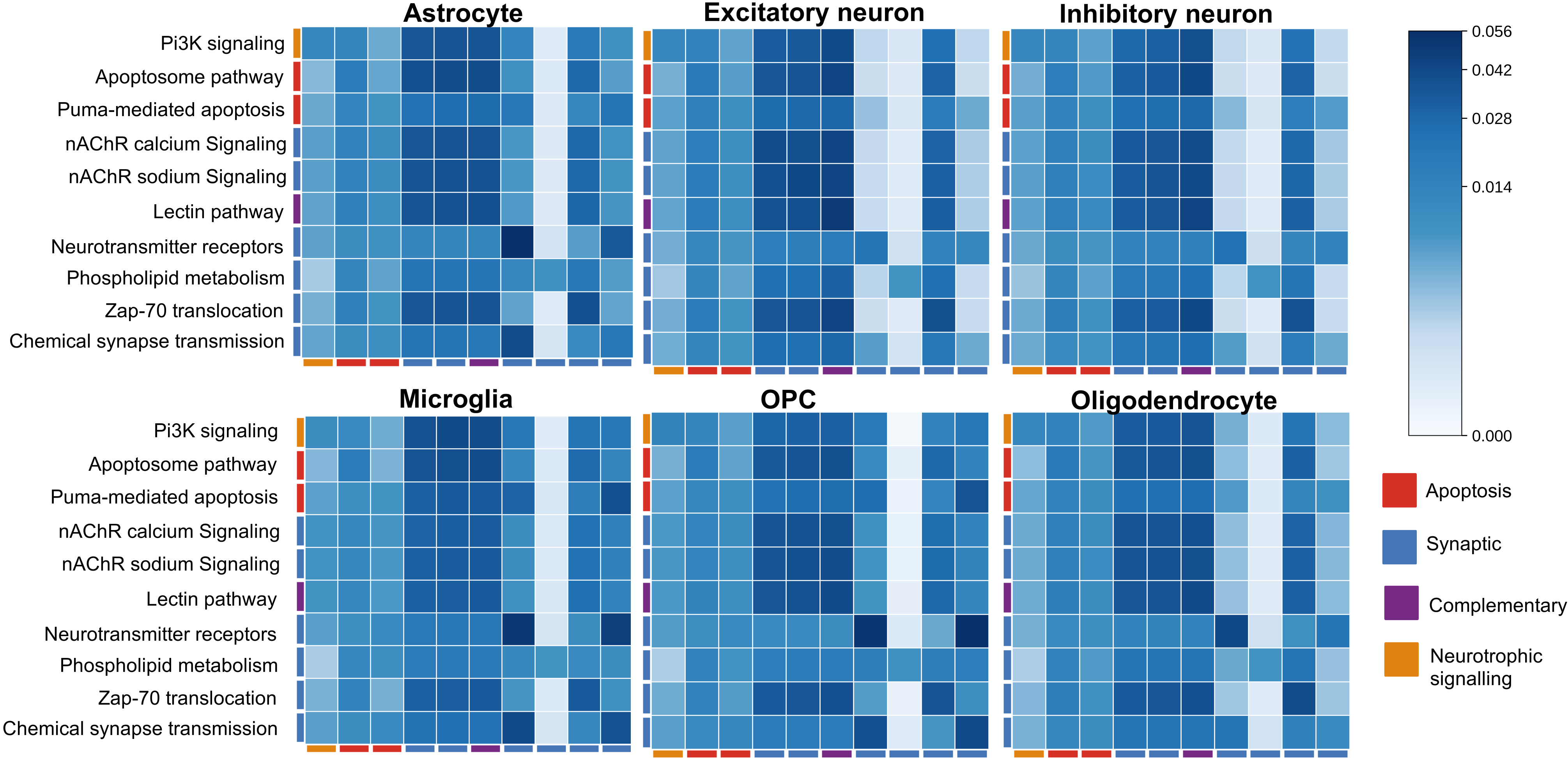}
\caption{
\textbf{Per-cell-type attention heatmaps, Reactome vocabulary. Mean inter-program attention for the top-variance Reactome programs, stratified by cell type.}
}
\label{fig:reactome_pathway_activation}
\end{figure}

The main-text attention heatmaps (Fig.~\ref{fig:healthy_heatmaps}) pool $\bar{A}$ across all nuclei. Decomposing $\bar{A}$ by cell type tests a stronger claim: that ORBIT recovers \emph{cell-type-specific} co-activation structure rather than a single tissue-wide template reweighted by composition. Figures~\ref{fig:allen_pathway_activation}--\ref{fig:reactome_pathway_activation} show, for each vocabulary, the head-averaged attention matrix $\bar{A}_{ct}$ restricted to a single cell type $ct$, computed on control nuclei. Rows are a fixed set of high-variance programs across all six panels of a given figure, allowing direct comparison of how a single program's attention reorganizes across cell types.

\textbf{Allen Brain Atlas.}
ABA shows the strongest cell-type stratification of the three vocabularies, consistent with ABA programs being themselves cell-type-defined. Within astrocytes, Astro L1 FGFR3 SERPINI2 and Astro L1-6 FGFR3 PLCG1 concentrate outgoing attention on a narrow band of astrocyte targets, reproducing the within-lineage coupling reported in Section~3.4. Excitatory and inhibitory panels show the complementary pattern: FEZF2 and PVALB programs (Exc L5 FEZF2 PKD2L1, Inh L1-2 PVALB CDK20) become the dominant attention drivers, while Exc L3 THEMIS ENPEP and Inh L5-6 SST DNAJC14 act as integrators. Microglia exhibit a markedly compressed dynamic range relative to neurons or astrocytes, consistent with their smaller and more homogeneous program repertoire in healthy cortex. OPC and oligodendrocyte panels show broader attention coupling across nearly all column targets, indicating that cells of the oligodendrocyte lineage maintain stronger cross-program integration than terminally differentiated neurons in this vocabulary.

\textbf{KEGG.}
KEGG programs are organized by signaling cascade rather than cell identity, so cell-type-specific attention is expressed through which cascade modules co-activate. The synaptic block (eCB signaling, glutamatergic synapse, long term potentiation (LTP), calcium signaling, synaptic vesicle cycle) forms a tightly coupled cluster across all six cell types, with absolute attention magnitude highest in OPC and oligodendrocyte panels and most attenuated in microglia. The Alzheimer's disease and Parkinson's disease programs cluster with each other but receive only weak attention from the synaptic block, consistent with the gene-level finding (Section~3.4) that hub kinases (MAPK1, AKT1/3, PIK3CA) route influence through synaptic programs rather than through the named neurodegeneration pathways. This separation is preserved across all six cell types, indicating that vocabulary-level routing is not a cell-type artifact.

\textbf{Reactome.}
Reactome vocabulary ($P=170$) holds the same block structure: chemical synapse transmission, neurotransmitter receptors, and the nicotinic acetylcholine receptors (nAChR) calcium and sodium signaling programs form a coupled synaptic block, while Apoptosome pathway and PUMA-mediated apoptosis form a smaller separate cluster. The synaptic block is most prominent in excitatory and inhibitory panels and gains relative weight against apoptosis programs in oligodendrocyte and OPC panels. Cross-vocabulary alignment is non-trivial: chemical synapse transmission (Reactome) and synaptic vesicle cycle (KEGG) recover the same neuron-enriched attention pattern despite distinct gene sets, indicating that the cell-type structure ORBIT extracts is a property of the underlying transcriptome rather than of any particular pathway annotation. 
Note that the absolute attention magnitudes in Reactome (max $\approx 0.056$) are systematically lower than in ABA ($\approx 0.13$) or KEGG ($\approx 0.12$). This is likely a consequence of softmax peak height, which depends on the separability of program embeddings rather than their count. Reactome's hierarchical organization places pathways with substantial shared membership in the same vocabulary (chemical synapse transmission $\supset$ neurotransmitter receptors $\supset$ nAChR signaling), so near-identical masked inputs produce near-identical embeddings and softmax mass splits across clusters of indistinguishable keys. Relative structure within $\bar{A}_{\text{Reactome}}$, including block organization, cell-type contrasts, the synaptic-vs-apoptosis partition, is preserved and remains the interpretable quantity.

\textbf{Biological Implications}
The fact that the same trained model produces qualitatively distinct $\bar{A}_{ct}$ matrices across cell types without any cell-type label entering Stage~1 supports the central claim that the attention matrix encodes biological organization recoverable from expression covariance alone.

\subsection{Gene Influence Propagation in Healthy Tissue}
\begin{figure}[t]
\centering
\includegraphics[width=\linewidth]{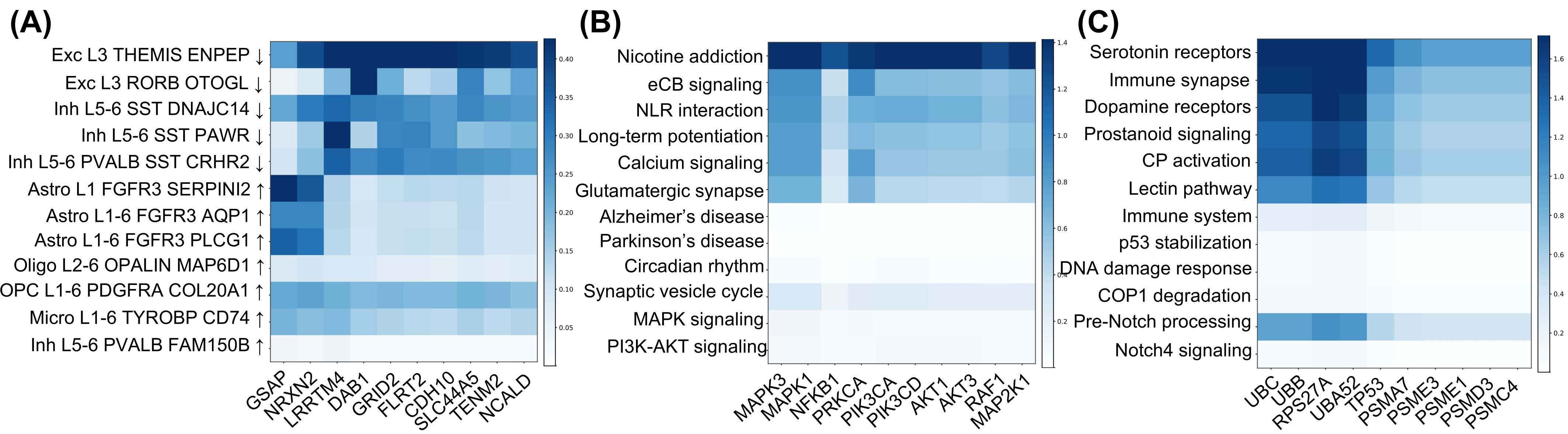}
\caption{
\textbf{ORBIT propagates gene influence through the attention graph.} \textbf{a}, Pathway influence heatmap for the top 10 data-adaptively identified influence genes (ABA). \textbf{b}, Pathway influence heatmap for top 10 influence genes (KEGG). \textbf{c}, Pathway influence heatmap for top 10 influence genes (Reactome).
}
\label{fig:gene_to_pathway}
\end{figure}

Figure~\ref{fig:gene_to_pathway} reports the gene-pathway influence matrix $I_{g,p_2}$ defined in Eq.~\ref{eq:influence}, restricted to the top 10 highest-influence genes per vocabulary. The matrix combines the binary gene-program membership mask $M$ with the learned attention weights $\bar{A}$: a high $I_{g,p_2}$ indicates that gene $g$ belongs to one or more programs $p_1$ that, on average, place high attention on $p_2$. The matrix is computed entirely post-hoc with no additional learning, and gene rankings come from the attention weights alone, since membership $M$ assigns each gene a constant total mass of $|\{p_1 : M_{g,p_1}=1\}|$.

\textbf{ABA (Fig.~\ref{fig:gene_to_pathway}a).}
The top-10 ABA influence genes split into two visually distinct clusters. \textit{GSAP} and \textit{NRXN2} show their highest influence on the three astrocyte FGFR3 programs (SERPINI2, AQP1, PLCG1), visible as the dark band in the upper-left corner. \textit{GSAP} encodes $\gamma$-secretase activating protein, which directly promotes APP cleavage~\citep{jin2022}; routing into a glial program rather than a neuronal one is non-trivial because no AD or cell-type label was supplied during Stage~1. The remaining genes (\textit{LRRTM4}, \textit{DAB1}, \textit{GRID2}, \textit{FLRT2}, \textit{CDH10}, \textit{SLC44A5}, \textit{TENM2}, \textit{NCALD}) are synaptic organizers and adhesion molecules, routing influence primarily into the excitatory and interneuron integrator programs (Exc L3 THEMIS ENPEP, Inh L5-6 SST DNAJC14, Inh L5-6 SST PAWR). Microglial (TYROBP CD74) and OPC programs receive weak influence from this gene set, consistent with the genes' neuronal expression bias in healthy cortex.

\textbf{KEGG (Fig.~\ref{fig:gene_to_pathway}b).}
The KEGG top-10 list is uniformly dominated by hub kinases of the MAPK/PI3K/AKT cascade (\textit{MAPK3}, \textit{MAPK1}, \textit{NFKB1}, \textit{PRKCA}, \textit{PIK3CA}, \textit{PIK3CD}, \textit{AKT1}, \textit{AKT3}, \textit{RAF1}, \textit{MAP2K1}), none of which are canonical AD or PD genes. Their influence concentrates in synaptic and addiction programs (Nicotine addiction, eCB signaling, LTP, calcium signaling, glutamatergic synapse) and on the immune NLR interaction program.

\textbf{Reactome (Fig.~\ref{fig:gene_to_pathway}c).}
The Reactome top-10 is dominated by ubiquitin-proteostasis machinery: the ubiquitin-ribosomal fusion proteins \textit{UBC}, \textit{UBB}, \textit{RPS27A}, \textit{UBA52}, the tumor suppressor \textit{TP53}, and five proteasome subunits (\textit{PSMA7}, \textit{PSME3}, \textit{PSME1}, \textit{PSMD3}, \textit{PSMC4}). Influence concentrates in the upper rows: serotonin receptors, immune synapse, dopamine receptors, prostanoid signaling, complement (CP) activation, and lectin pathway. Influence on p53 stabilization, DNA damage response, COP1 degradation, and Notch processing is comparatively low, even though \textit{TP53} is a member of those programs by annotation. The ordering reflects attention magnitude rather than membership count, indicating that ORBIT distinguishes the programs through which the proteostasis machinery genuinely co-activates from those in which membership is merely formal.

\subsection{Extended Per-Cell-Type Classification Metrics}
\label{app:classification}
Table~\ref{tab:per_class_f1} extends the macro-F1 comparison from Table~\ref{tab:classification} to per-class F1 across all six major cell types (vascular cells are oversampled and excluded from macro averaging, as described in Section~3.1). ORBIT's compression to 220 pathway scores does not preferentially fail on any single cell population. ORBIT achieves its highest per-class F1 on astrocytes (0.992), exceeding both CellTypist (0.989) and the MLP baseline (0.984), consistent with the strong astrocyte-specific co-activation structure visible in Fig.~\ref{fig:allen_pathway_activation} (Astrocyte panel). On excitatory neurons, ORBIT (0.976) is within 0.003 of CellTypist (0.979) and 0.004 of the MLP, reflecting the inherent difficulty of distinguishing fine excitatory subtypes from a 220-dimensional input. The 0.014 macro-F1 gap to scANVI reflects scANVI's use of an explicit reference-mapping prior, which is complementary to ORBIT rather than in competition: ORBIT's primary output is the pathway attention matrix, and the classification head is a thin auxiliary used to enable condition-stratified analysis.

\begin{table}[h]
\centering
\caption{Per-class F1 scores on Morabito 2021 (80/20 stratified split, 
seed 42). Vasc: vascular cells (oversampled; excluded from macro average). 
Oli: oligodendrocytes; OPC: oligodendrocyte precursor cells.}
\label{tab:per_class_f1}
\small
\begin{tabular}{lccccccc}
\toprule
Method & Exc & Inh & Ast & Mic & Oli & OPC & Macro F1 \\
\midrule
CellTypist & 0.979 & 0.988 & 0.989 & 0.990 & 0.994 & 0.980 & 0.987 \\
scANVI & 0.990 & 1.000 & 0.990 & 1.000 & 1.000 & 1.000 & 0.990 \\
MLP Baseline & 0.980 & 0.984 & 0.984 & 0.986 & 0.994 & 0.994 & 0.979 \\
\textbf{ORBIT} & \textbf{0.976} & \textbf{0.982} & \textbf{0.992} 
  & \textbf{0.978} & \textbf{0.993} & \textbf{0.989} & \textbf{0.984} \\
\bottomrule
\end{tabular}
\end{table}

\subsection{AUCell Agreement Stratified by Program Size}
\label{app:aucell}
Table~\ref{tab:aucell} stratifies the Pearson agreement between ORBIT pathway scores and AUCell rAUC scores by program size. The two metrics agree strongly on small-to-medium programs and degrade for programs with more than 100 member genes. The degradation reflects AUCell's known score dilution: AUCell ranks all genes per cell and computes the area under the recovery curve restricted to the top 5\% of ranked genes by default, so a 200-gene program competes for a fixed-size rank window with smaller programs and its rAUC saturates near a constant. 

ORBIT's score (Eq.~\ref{eq:score}) uses size-normalized summation, preserving cross-cell variance for large programs without rank truncation. Two observations support this interpretation. First, ORBIT scores for large programs remain internally consistent: their Pearson correlation with the same program's score on a held-out cohort is preserved at the same level as for small programs (Spearman $r_s = 0.642$ across the full attention matrix; Section~3.6). Second, the cross-vocabulary classification result (Table~\ref{tab:classification}, 0.984 macro F1) is achieved using the full vocabulary including large programs, indicating that large-program scores carry cell-type-discriminative information AUCell discards. The size-stratified disagreement therefore reflects ORBIT's pathway encoder capturing information AUCell cannot, rather than a failure of the encoder.

\begin{table}[h]
\centering
\caption{Pearson correlation between ORBIT pathway scores and AUCell rAUC scores, stratified by program size (number of member genes), ABA vocabulary.}
\label{tab:aucell}
\small
\begin{tabular}{lcc}
\toprule
Program size bin & Mean Pearson $r$ & Fraction $r > 0.9$ \\
\midrule
$\leq 25$ genes  & 0.87 & 64.2\% \\
26--50 genes     & 0.82 & 57.1\% \\
51--100 genes    & 0.72 & 41.8\% \\
${>}100$ genes   & 0.28 & 12.3\% \\
\midrule
All (220 programs) & 0.922 (median) & 52.3\% \\
\bottomrule
\end{tabular}
\vspace{3pt}
\footnotesize

Degradation at large program sizes reflects AUCell's known score dilution for large gene sets, not a failure of ORBIT. ORBIT scores for large programs remain internally consistent across nuclei.
\end{table}

\clearpage
\section*{NeurIPS Paper Checklist}

\begin{enumerate}

\item {\bf Claims}
    \item[] Question: Do the main claims made in the abstract and introduction accurately reflect the paper's contributions and scope?
    \item[] Answer: \answerYes{}
    \item[] Justification: The abstract and introduction consistently state that ORBIT learns a directed program-program dependency structure from observational single-cell data via an intervention-consistent objective. These claims match the experimental results (synthetic recovery, AD rewiring patterns, and classification performance) and the stated scope limitations (representation-space directionality, not causal inference).

\item {\bf Limitations}
    \item[] Question: Does the paper discuss the limitations of the work performed by the authors?
    \item[] Answer: \answerYes{}
    \item[] Justification: The Discussion section explicitly addresses multiple limitations. Computational and generalization limitations across datasets and vocabularies are also discussed.

\item {\bf Theory assumptions and proofs}
    \item[] Question: For each theoretical result, does the paper provide the full set of assumptions and a complete (and correct) proof?
    \item[] Answer: \answerNA{}
    \item[] Justification: ORBIT does not present formal theorems or proofs. Its contributions are algorithmic and empirical: the intervention-consistent influence loss, the capacity-restricted projection $\varphi$, and the two-stage training framework are motivated conceptually and validated empirically across three pathway vocabularies and a synthetic dataset. Assumptions regarding fixed pathway vocabulary and zero-intervention approximation are stated explicitly in the Methods and Discussion sections.

\item {\bf Experimental result reproducibility}
    \item[] Question: Does the paper fully disclose all the information needed to reproduce the main experimental results of the paper to the extent that it affects the main claims and/or conclusions of the paper?
    \item[] Answer: \answerYes{}
    \item[] Justification: The manuscript specifies dataset sources (with GEO accession IDs), preprocessing (log1p normalization, library size normalization), model architecture (layer dimensions, heads, dropout), training schedules (optimizer, learning rate, warmup, batch size, epochs), masking ratios, intervention loss strategy, and evaluation protocols (train/test splits, permutation tests, and vocabulary definitions). These details are sufficient to reproduce the core experimental findings and ablation structure.

\item {\bf Open access to data and code}
    \item[] Question: Does the paper provide open access to the data and code, with sufficient instructions to faithfully reproduce the main experimental results?
    \item[] Answer: \answerYes{}
    \item[] Justification: The paper links to an anonymized GitHub repository in the abstract, and the repository README provides installation instructions, runnable commands, and benchmark scripts. The datasets used are retrieved from the Gene Expression Omnibus public repository.

\item {\bf Experimental setting/details}
    \item[] Question: Does the paper specify all the training and test details necessary to understand the results?
    \item[] Answer: \answerYes{}
    \item[] Justification: The paper provides detailed descriptions of data sources, preprocessing pipelines, model architecture (dimensions, attention heads, dropout), optimization settings (AdamW, learning rate schedules, warmup ratios, gradient clipping), training splits (80/20 stratified, held-out cohorts), and evaluation procedures (macro F1, permutation testing, AUCell comparisons). Hyperparameter selection procedure and vocabulary construction are also described.

\item {\bf Experiment statistical significance}
    \item[] Question: Does the paper report error bars or statistical significance appropriately?
    \item[] Answer: \answerYes{}
    \item[] Justification: The paper uses permutation testing (1,000 permutations) with Benjamini--Hochberg FDR correction to assess significance of rewiring edges, reports stability across subsamples (Jaccard stability of top pairs), and compares observed attention shifts against null distributions. Cross-dataset correlations and classification performance are reported on fixed splits. Classification performance is evaluated on a fixed stratified split; no confidence intervals are reported for F1 as a single split is used, consistent with practice in the single-cell classification literature.

\item {\bf Experiments compute resources}
    \item[] Question: Does the paper provide sufficient information on compute resources?
    \item[] Answer: \answerYes{}
    \item[] Justification: The compute resources used for all experiments are described in the appendix.

\item {\bf Code of ethics}
    \item[] Question: Does the research conform with the NeurIPS Code of Ethics?
    \item[] Answer: \answerYes{}
    \item[] Justification: The study uses publicly available de-identified post-mortem transcriptomic datasets and does not involve human experimentation, intervention, or re-identification risks. The methodology does not introduce foreseeable ethical violations under NeurIPS guidelines. Limitations and need for biological validation are explicitly acknowledged.

\item {\bf Broader impacts}
    \item[] Question: Does the paper discuss both positive and negative societal impacts?
    \item[] Answer: \answerNA{}
    \item[] Justification: ORBIT is a computational method for analyzing gene program co-activation structure in single-cell transcriptomic data. Its primary application domain is basic neuroscience and disease biology research. We do not foresee direct positive or negative societal impacts from this work beyond those inherent to any basic research tool.

\item {\bf Safeguards}
    \item[] Question: Does the paper describe safeguards for responsible release?
    \item[] Answer: \answerNA{}
    \item[] Justification: The work does not release high-risk models or datasets requiring safety gating. It uses publicly available transcriptomic data and a research-grade model intended for scientific analysis rather than deployment in sensitive or high-risk applications.

\item {\bf Licenses for existing assets}
    \item[] Question: Are existing assets properly credited and licensed?
    \item[] Answer: \answerYes{}
    \item[] Justification: All datasets (e.g., GSE174367, KEGG, Reactome, Allen Brain Atlas, AUCell references) are properly cited with corresponding literature references. Standard public bioinformatics resources are used in accordance with their published usage policies.

\item {\bf New assets}
    \item[] Question: Are new assets introduced in the paper well documented and is the documentation provided alongside the assets?
    \item[] Answer: \answerYes{} 
    \item[] Justification: The paper releases new code via an anonymized GitHub repository, and the repository includes documentation with installation instructions, runnable commands, benchmark scripts, and a description of the repository layout.

\item {\bf Crowdsourcing and research with human subjects}
    \item[] Question: For crowdsourcing experiments and research with human subjects, does the paper include the full text of instructions given to participants and screenshots, if applicable, as well as details about compensation (if any)? 
    \item[] Answer: \answerNA{} 
    \item[] Justification: The paper does not involve crowdsourcing experiments or research with human subjects.

\item {\bf Institutional review board (IRB) approvals or equivalent for research with human subjects}
    \item[] Question: Does the paper describe potential risks incurred by study participants, whether such risks were disclosed to the subjects, and whether Institutional Review Board (IRB) approvals (or an equivalent approval/review based on the requirements of your country or institution) were obtained?
    \item[] Answer: \answerNA{} 
    \item[] Justification: The paper does not involve research with human subjects.

\item {\bf Declaration of LLM usage}
    \item[] Question: Does the paper describe the usage of LLMs if it is an important, original, or non-standard component of the core methods in this research? Note that if the LLM is used only for writing, editing, or formatting purposes and does \emph{not} impact the core methodology, scientific rigor, or originality of the research, declaration is not required.
    \item[] Answer: \answerNA{} 
    \item[] Justification: The core method, experiments, and evaluation do not use LLMs as an important, original, or non-standard component of the research.
    
\end{enumerate}
\end{document}